\documentclass[reqno]{amsart}
\usepackage{amsmath, amsfonts, amsthm}

\newtheorem*{hh}{Theorem}
\newtheorem*{pr}{Proposition}

\theoremstyle{remark}
\newtheorem*{remark}{Remark}
\newtheorem*{remarks}{Remarks}
\newcommand{\N}{{\mathbb N}}

\newcommand{\Z}{{\mathbb Z}}

\title[Event Horizon Spectrum]{Intrinsic Spectral Geometry of the Kerr-Newman Event Horizon\\}
\author{Martin Engman}
\author{Ricardo Cordero Soto}
\begin{document}
\address{Departamento de Ciencias y Tecnolog\'{i}a, Universidad Metropolitana, 
San Juan, PR 00928}
\email{um\_mengman@suagm.edu}
\email{mathengman@yahoo.com}

\begin{abstract}
{We uniquely and explicitly reconstruct the instantaneous intrinsic metric of the Kerr-Newman Event Horizon from the 
spectrum of its 
Laplacian. In the process we find that the angular momentum parameter, radius, area; and in the uncharged case, mass, 
can be written 
in terms of these eigenvalues. In the uncharged case this immediately leads to the unique and explicit
 determination 
of the Kerr metric in terms of the spectrum of the event horizon. Robinson's ``no hair" theorem now yields the corollary:  One can ``hear the shape" of noncharged stationary axially symmetric black hole space-times by listening to the vibrational frequencies of its event horizon only.} 
\end{abstract}
\maketitle

\section{Introduction}

Although in general the spectrum of the Laplacian on a manifold determines, via the heat kernel, a sequence of invariants 
which restrict the geometry, it is a rare occasion indeed that the metric is uniquely determined by the eigenvalues (see, 
for example, 
Gordon \cite{gor}). In fact one of the few uniqueness results of this kind was proved by Br\"{u}ning and Heintze, \cite{bh}: An
 $S^1$ invariant 
two dimensional surface diffeomorphic to the sphere, which has, in addition, a mirror symmetry about its equator is uniquely 
determined by its spectrum. This has 
been generalized by Zelditch, \cite{zeld}, but the Br\"{u}ning and Heintze result is the appropriate setting for our purposes since 
this is exactly the class of metrics which include Kerr-Newman event horizons. 

In 1973 Smarr \cite{smarr} studied the metric of the horizon in terms of a scale parameter and a distortion parameter and in
a particularly convenient coordinate system for calculating, for example, the curvature. More recently, the first author of 
the present paper has used a similar coordinate system to study the spectrum of $S^1$ invariant surfaces. As it turns out, 
Smarr's form of the horizon metric is a scaled version of a special case of our form of the metric.

$S^1$ invariant metrics on $S^2$ can be described in terms of a single function $f(x)$. One can show that the sum of the 
reciprocals of the nonzero, $S^1$ invariant eigenvalues (i.e. the trace of the $S^1$ invariant Greens operator) is given by 
an integral involving the function $f(x)$. The key to 
the results we obtain here is that, in the case of the Kerr-Newman event horizon, the function $f(x)$ is a simple
rational function and the above mentioned integral can be calculated explicitly. Together with a similar equivariant trace 
formula, one can now write Smarr's parameters in terms of these traces and hence explictly display the metric in terms 
of it's spectrum. As far as we know, this is the only nontrivial example of this phenomonen (the Br\"{u}ning and Heintze 
result is not constructive in this sense).

An interesting byproduct of our results (in the uncharged case), together with Robinson's uniqueness theorem, is the unique reconstruction of space-time in terms of these eigenvalues. We conclude the paper with a discussion of possible physical interpretations of these results.

\section{Smarr's form of the metric}

The {\em Kerr-Newman metric} describing the geometry of a rotating charged black hole written in Kerr ingoing coordinates 
$(v,r, \theta, \phi)$ is

\begin{eqnarray}
\lefteqn{ds^2=-\left(1-\frac{2mr-e^2}{\Sigma}\right)dv^2+2drdv-\frac{(2mr-e^2)2a\sin^2\theta}{\Sigma}dvd \phi} 
\hspace{3.7in} \label{eq:kerr}
\\
 -2a\sin^2\theta drd \phi+\Sigma d
\theta^2+\left(\frac{(r^2+a^2)^2-\delta a^2\sin^2\theta}{\Sigma}\right)\sin^2\theta d \phi^2, \nonumber
\end{eqnarray}
in which $(m,a,e)$ represent respectively, the total mass, angular momentum per unit mass, and the charge. Also 
$\Sigma=r^2+a^2\cos^2\theta$ and $\delta=r^2-2mr+a^2+e^2$ (we use the lower case letter here, instead of the 
traditional upper case, to avoid confusion with the equally traditional use of $\Delta$ for the Laplacian). The uncharged 
($e=0$) case is the {\em Kerr metric}.

This family of metrics is quite general. To quote Wald \cite{wal} ``. . .  the Kerr-Newman family of solutions completely describes all the stationary black holes which can possibly occur in general relativity." This fact is due to the 
famous uniquess theorems of Israel \cite{isr},  Carter and Robinson \cite{rob}, and Mazur \cite{maz} for example.  

For the Kerr-Newman metric the surface of the event horizon can be thought of as a spacelike slice through the null 
hypersurface defined by
 the largest root, $r_+$, of $\delta = 0$ i.e. $r_+=m+\sqrt{m^2-a^2-e^2}$. 
The intrinsic instantaneous metric on the event horizon is obtained by setting $r=r_+$, so that $dr=0$, and also 
setting $dv=0$ to get
\begin{equation}
 ds_{eh}^2=(r_+^2+a^2\cos^2\theta)d\theta^2+\left(\frac{(r_+^2+a^2)^2}{r_+^2+a^2\cos^2\theta}\right)\sin^2\theta d \phi^2 
\label{eq:surfkerr}
\end{equation}

In \cite{smarr}, Smarr defines {\em the scale parameter} by $\eta=\sqrt{r_+^2+a^2}$ and {\em the distortion parameter} by 
$\beta = \frac{a}{\sqrt{r_+^2+a^2}}$ and introducing a new variable $x=\cos \theta$ one finds that the event horizon 
metric is 

\begin{equation}
 ds_{eh}^2=\eta^2\left(\frac{1}{f(x)}dx^2 + f(x)d\phi^2\right)  \label{eq:xeh}
\end{equation}
where $(x,\phi) \in(-1,1) \times [0,2\pi)$ and 
\begin{equation} 
f(x)=\frac{1-x^2}{1-\beta^2(1-x^2)}. \label{eq:function}
\end{equation} It is well known that the Gauss 
curvature of a metric in this form is simply $K(x)=-1/(2\eta^2)f''(x)$ so that in this case
\begin{equation}
K(x)=\frac{1}{\eta^2}\left(\frac{1-\beta^2(1+3x^2)}{(1-\beta^2(1-x^2))^3}\right), \label{eq:curv}
\end{equation}
and the surface area of the metric is $A=4\pi \eta^2$.
We point out that in case $a=0$ ($\beta = 0$), $e=0$  \eqref{eq:kerr} gives the Schwarzschild black hole and \eqref{eq:xeh} 
is the metric of the constant Gauss curvature $=\frac{1}{\eta^2}$ metric on $S^2$.  

\section{Spectrum of $S^1$ invariant metrics}

For any Riemannian manifold with metric $g_{ij}$ {\em the Laplacian} is given by

$$\Delta_{g} =- \frac{1}{\sqrt{g}} \frac{\partial }{\partial x^i} \left( \sqrt{g}g^{ij} 
\frac{\partial }{\partial x^j} \right) . $$

This is the Riemannian version of the {\em Klein-Gordon}, or {\em D'Alembertian}, or 
{\em wave} operator usually denoted by $\Box$. 

In this section we outline some our previous work on the spectrum of the Laplacian on $S^1$ invariant metrics on $S^2$. 
The interested reader may consult \cite{eng1}, \cite{eng2} and 
\cite{eng3} for further details. 

To simplify the discussion the area of the metric is normalized to $A=4\pi$ for this section only.
The metrics we study have the form:
\begin{equation}
dl^2=\frac{1}{f(x)}dx^2 + f(x)d\phi^2  \label{eq:surfrev}
\end{equation}
where $(x,\phi) \in(-1,1) \times [0,2\pi)$ and $f(x)$ satisfies $f(-1)=0=f(1)$ and 
$f'(-1)=2=-f'(1)$. 
In this form, it is easy to see that the 
Gauss curvature of this metric is given by 
$K(x) =(-1/2)f^{''}(x)$.
 The canonical (i.e. constant curvature) metric is obtained by taking 
$f(x)=1-x^{2}$ and the metric \eqref{eq:xeh} is a homothety (scaling) of a particular example of the general form 
\eqref{eq:surfrev}. 

The Laplacian for the metric \eqref{eq:surfrev} is

$$\Delta_{dl^2} =  -\frac{\partial }{\partial x} \left( f(x) 
\frac{\partial }{\partial x} \right) - \frac{1}{f(x)}\frac{\partial^2 }{\partial \phi^2}.$$
Let $\lambda$ be any eigenvalue of $-\Delta$.
We will use the symbols $E_{\lambda}$ and $\dim E_{\lambda}$ to denote the 
eigenspace for $\lambda $ and it's multiplicity (degeneracy) respectively. In this paper 
the symbol $\lambda_m$ will always mean the $m$th \underline{distinct} 
eigenvalue. We adopt the convention $\lambda_0=0$. 
Since $S^1$ (parametrized here by $0 \leq \phi <2\pi $) acts
 on
$(M,g)$ by isometries we can separate variables and because $\dim E_{\lambda_m} \leq 2m+1$ (see 
\cite{eng3} for the proof), the 
orthogonal decomposition of $E_{\lambda_m}$ has the special form

\begin{equation*}
E_{\lambda_m}= \bigoplus_{k=-m}^{k=m} e^{ik\theta}W_k
\end{equation*}
in which $W_k (=W_{-k})$ is the ``eigenspace" (it might contain only $0$)
 of the ordinary differential 
operator

\begin{equation*}
L_{k}=-\frac{d}{dx}\left(f(x)\frac{d}{dx}\right) + \frac{k^2}{f(x)}
\end{equation*}
with suitable boundary conditions. It should be observed that $\dim W_k \leq
1$, a value of zero for this dimension occuring when $\lambda_m$ is not in the spectrum of
$L_k$.

The set of positive eigenvalues is given by $Spec(dl^2)= \bigcup_{k\in \Z} Spec L_k$ 
and consequently the nonzero part of the
spectrum of $-\Delta$ can be studied via the spectra 
$Spec L_{k}=\{0<\lambda_{k}^{1} < \lambda_{k}^{2} < \cdots 
<\lambda_{k}^{j} < \cdots \} \forall k \in \Z$. 
The eigenvalues $\lambda^j_0$ in the case $k=0$ above are called the 
{\em $S^1$ invariant eigenvalues} since their eigenfunctions are 
invariant under the action of the $S^1$ isometry group. If $k \neq 0$ the eigenvalues
are called {\em $k$ equivariant} or simply {\em of type $k \neq 0
$}. Each $L_{k}$ has a Green's operator, 
$\Gamma_{k}:(H^{0}(M))^{\perp} \rightarrow L^{2}(M)$, whose spectrum 
is $\{ 1/\lambda_{k}^{j} \}_{j=1}^{\infty}$, and whose trace is defined by 
\begin{equation}
\gamma_{k} \equiv \sum_{j=1}^{\infty} \frac{1}{\lambda_{k}^{j}}. \label{eq:trace}
\end{equation}

The formulas of present interest were derived in \cite{eng1} and \cite{eng2} and are given by  

\begin{equation}
 \gamma_{0}=
   \frac{1}{2} \int_{-1}^{1} \frac{1-x^{2}}{f(x)} dx \label{eq:gam0}
\end{equation}
and
 \begin{equation}  
   \gamma_{k}=\frac{1}{|k|} \hspace{.2in}  \mbox{if $k \neq 0$} \label{eq:gamk}
\end{equation}

\begin{remark}

 One must be careful with the definition of $\gamma_0$ since 
$\lambda^0_0 = 0$ is an $S^1$ invariant eigenvalue of $-\Delta$. To avoid this difficulty we studied the $S^1$ invariant 
spectrum
of the Laplacian on 1-forms in \cite{eng2} and then observed that the 
nonzero eigenvalues are the same for functions and 1-forms.

\end{remark}

\section{Spectral Determination of the Event Horizon}

In case $f(x)$ is given by \eqref{eq:function} the metric \eqref{eq:xeh} is related to \eqref{eq:surfrev} via the homothety
$ds_{eh}^2=\eta^2dl^2$, and it is well known that 
$$\lambda \in Spec(dl^2) \hspace{.2in} \mbox{if and only if} \hspace{.2in} \frac{\lambda}{\eta^2} \in Spec(\eta^2 dl^2)$$ 
so that, after an elementary integration, the trace formulae for the event horizon are

\begin{equation}
 \gamma_{0}=\eta^2\left(1-\frac{2\beta^2}{3}\right) \label{eq:ehgam0}
\end{equation}
and
 \begin{equation}  
   \gamma_{k}=\frac{\eta^2}{|k|} \hspace{.2in}  \forall k \neq 0 \label{eq:ehgamk}
\end{equation}

For the $k=1$ case, for example, one can invert the resulting pair of equations to get

\begin{equation}
 \beta^2=\frac{3}{2}\left(1-\frac{\gamma_{0}}{\gamma_{1}}\right) \label{eq:invbeta}
\end{equation}
and
 \begin{equation}  
 \eta^2= \gamma_{1}, \label{eq:inveta}
\end{equation}
and thereby write the event horizon metric explicitly and uniquely in terms of the spectrum as follows.

\begin{pr} With $\gamma_0$ and $\gamma_1$ defined as in \eqref{eq:trace} the instantaneous intrinsic metric of the Kerr-Newman event horizon is given by
\begin{equation}
 ds_{eh}^2=\gamma_1\left(\frac{1-\frac{3}{2}\left(1-\frac{\gamma_{0}}{\gamma_{1}}\right)(1-x^2)}{1-x^2}dx^2 + 
\frac{1-x^2}{1-\frac{3}{2}\left(1-\frac{\gamma_{0}}{\gamma_{1}}\right)(1-x^2)}d\phi^2\right).  \label{eq:finalmetric}
\end{equation}
\end{pr}

An immediate consequence of \eqref{eq:ehgamk} is that the area of the metric has a representation for each $k \in \N$ given
by
\begin{equation}
A=4\pi k\gamma_k. \label{eq:area}
\end{equation}

\begin{remarks} 

i.) One might view equations Eqs. \eqref{eq:ehgam0} and \eqref{eq:ehgamk} as nothing more than 
definitions of new parameters in terms of the old. On the other hand, the quantities $\gamma_k$ are fundamental 
and naturally defined quantities coming from the discrete set of vibrational wave frequencies on the surface. Alternatively, one 
can think of each trace, $\gamma_k$ as the sum of the squares of all wavelengths of given quantum number $k$.

ii.) One can use $\gamma_0$ together with $\gamma_k$ for any $k$ to reconstruct the metric. In some sense one can ``construct the 
metric in $\aleph_0$ ways".

iii.) Br\"uning and Heintze proved that for $S^1$ invariant metrics symmetric about the equator the $S^1$ invariant 
spectrum determines the metric. Their result requires the prescription of all of the eigenvalues of the $k=0$ spectrum to 
uniquely determine the surface of revolution. In the 
example of the present paper the metric is, therefore, uniquely determined once we have knowledge of the entire list of 
$S^1$ invariant eigenvalues. For the explicit construction one exchanges complete knowledge of the $k=0$ spectrum for 
partial spectral data, namely the traces of the Greens operators for $k=0$ and any $k \neq 0$.
\end{remarks}

All the physical parameters can also be written in terms of the spectra as follows:

\begin{equation}
a^2=\frac{3}{2}(\gamma_1-\gamma_0) \label{eq:a}
\end{equation}
\begin{equation}
r_+^2=\frac{3\gamma_0-\gamma_1}{2} \label{eq:r}
\end{equation}
\begin{equation}
m^2=\frac{(\gamma_1+e^2)^2}{6\gamma_0-2\gamma_1}, \label{eq:m}
\end{equation}
as well as the angular velocity and surface gravity. 
We observe, however, that the mass depends on the charge $e$ as well as the spectrum so that the mass is uniquely 
determined by the spectrum only in the Kerr ($e=0$) case.

We set together Remark iii. with \eqref{eq:a}, \eqref{eq:m} (with $e=0$), \eqref{eq:kerr}, 
and Robinson's uniqueness Theorem \cite{rob} to observe. 

\begin{hh} A noncharged stationary axially symmetric asymptotically flat vacuum space-time with a regular event-horizon 
is uniquely determined by the 
$S^1$ invariant spectrum of the intrinsically defined Laplacian on its event horizon. The space-time is explicitly 
constructed from the $S^1$ invariant trace together with any $k \neq 0$ trace of the associated Green's operator on the 
event horizon. 
\end{hh}

\section{Discussion}
It is well known (see \cite{fn}, \cite{t}, \cite{n} among many others) that the quasinormal mode frequencies for scalar,
 electromagnetic, and gravitational radiation of black holes are related to the separation constants arising from
 separating variables in the Teukolsky master equation. These ``characteristic sounds" of the black hole are considered
 to be observable in an astrophysical sense. The angular operator (and, respectively, its separation constants) coming from
 the separation of variables is closely related 
to the Laplacian (and, respectively, its eigenvalues) on the event horizon in the sense that, for scalar fields, they both
 reduce to the Laplacian (and corresponding eigenvalues) on the constant curvature $S^2$ for the Schwarzschild ($a=0$)
 case.  There is also some evidence (this is a work in progress) that the $a>0$ Teukolsky angular equation 
can be viewed, after a change of variables, as a Laplace eigenvalue equation for a fourth order Taylor polynomial
 approximation of the horizon metric function. We hope to pursue these matters in our future work.

On a more philosophical note, the reader may have noticed that the main result of this paper is consistent with the holographic principle (\cite{sus}, \cite{th}) in as much as the structure of the ($3+1$ dimensional) Kerr-Newman space-time is encoded in the intrinsic spectral data of the (two-dimensional) event horizon surface.

These phenomena suggest that the spectrum of the Laplacian on event horizons is playing an important, if rather subtle and not well understood, role in the physics of the space-time manifold. 
  
\section{Acknowledgements} 
The authors thank Thomas Powell, Steve Zelditch, and Roberto Ram\'{\i}rez for their ideas and assistance in the 
preparation of this paper. This work was partially supported by the NSF Grant: Model Institutes for Excellence at UMET.

\end{document}